# Orbital Solutions and Absolute Elements of the W UMa Binary MW Pavonis

Gabriella E. Alvarez and James R. Sowell

*School of Physics, Georgia Institute of Technology, Atlanta, GA 30332*

`jim.sowell@physics.gatech.edu`

Richard M. Williamon

*Department of Physics, Emory University, Atlanta, GA 30322*

`phyrmw1@emory.edu`

Emilio Lapasset

*Observatorio Astronómico de Córdoba, 5000 Córdoba, Argentina*

`lapasset@oac.uncor.edu`

## ABSTRACT

We present differential $UBV$ photoelectric photometry obtained by Williamon of the short-period A-type W UMa binary MW Pav. With the Wilson-Devinney analysis program we obtained a simultaneous solution of these observations with the $UBV$ photometry of Lapasset (1977, 1980), the $V$ measurements by the $ASAS$ program (Pojmanski 2002), and the double-lined radial velocity measurements of Rucinski & Duerbeck (2006). Our solution indicates that MW Pav is in an overcontact state, where both components exceed their critical Roche lobes. We derive masses of $M_1 = 1.514 \pm 0.063\,M_\odot$ and $M_2 = 0.327 \pm 0.014\,M_\odot$, and equal-volume radii of $R_1 = 2.412 \pm 0.034\,R_\odot$ and $R_2 = 1.277 \pm 0.019\,R_\odot$ for the primary and secondary, respectively. The system is assumed to have a circular orbit and is seen at an inclination of $86.39° \pm 0.63°$. The effective temperature of the primary was held fixed at 6900 K, whereas the secondary's temperature was found to be $6969\pm10$ K. The asymmetry of the light curves requires a large, single star spot on the smaller, less massive secondary component. A consistent base solution, with different spot characteristics for the Williamon, Lapasset, and $ASAS$ data, was found. The modeled spot varied little during the 40-year range of photometric observations. The combined solution utilized a third light component and found that the period is changing at a rate of $dP/dt = (6.50 \pm 0.19) \times 10^{-10}$.

*Subject headings:* binaries: close — binaries: eclipsing — binaries: spectroscopic — stars: individual (MW Pav)

## 1. INTRODUCTION

MW Pav is an under-studied, southern, A-type W UMa binary system. These objects are low-mass systems in contact and often the surfaces have thermalized. Excellent articles on W UMa stars include those by Wilson (1978), Twigg (1979), and Van Hamme (1982a); the last study includes MW Pav in the analysis of evolutionary states of W UMa binaries. The light variability of MW Pav [HD 197070, SAO 257849, CD $-72°1636$, HIP 102508] was discovered by Eggen (1968) at the Mt. Stromlo and Siding Spring Observatories. His first determination of the period was 0.562979 days. Shortly thereafter, Williamon (1971) obtained a series of $UBV$ observations. He computed a revised period of 0.79499080 days but never performed a detailed solution of the system's characteristics. Lapasset (1977, 1980) obtained $UBV$ light curves a few years later and derived a solution by the Russell-Merrill method (Russell & Merrill 1952) and then later with the Wilson-Devinney software (Wilson & Devinney 1971) [hereafter referred to as WD]. Double-lined spectroscopic data were acquired by Rucinski & Duerbeck (2006) in 1998. The current *ASAS* program (Pojmanski 2002) regularly takes one $V$ measurement per night, and a full light curve has been obtained. These four sets are basically the complete collection of MW Pav observations. We have computed an orbital elements and absolute dimensions solution utilizing all of these.

## 2. OBSERVATIONS AND REDUCTIONS

Our first set of photometric observations were obtained by R.M. Williamon in 1970 using the 16-inch #1 reflector at the Cerro Tololo Inter-American Observatory. Standard $UBV$ filters were used with a refrigerated 1P21 photomultiplier to closely approximate the effective wavelength of the Johnson-Morgan passband system. The observations were recorded with a Honeywell strip-chart recorder, and deflections were read with a 5-second timing accuracy. All measurements of MW Pav were made differentially with respect to the comparison star SAO 257484 [CD $-72°1635$, CPD $-72°2550$], and these were corrected for atmospheric extinction by means of nightly coefficients determined from the comparison star via the technique of Hardie (1962). The heliocentric Julian dates and differential magnitudes for all 448 of the observations are given in Table 1.

The second set of differential $UBV$ photometry was obtained and described by Lapasset (1977, 1980). He acquired 401 measurements during the years 1972, 1974, and 1978 with the 1.54-m telescope of the Bosque Alegre Station of the Cordoba Observatory. His individual $UBV$ magnitudes as a function of HJD have never been published, and they are provided here in Table 2. [After the publication of his papers, the comparison star HD 197417 was determined to be a low amplitude Ap CrEu(Sr) (Houk & Cowley 1975) variable star, designated as V434 Pav, with a period of 4.55 days (Catalano & Renson 1988). We felt the observations obtained on HJD 2441589 in 1972 seemed to have been affected, so we omitted those 43 values in all three bandpasses in our calculations; however, these data are listed in Table 2 with all of the other Lapasset data.]

### 3. COMBINED LIGHT AND VELOCITY SOLUTION

A third set of photometry, which was obtained from the years 2000 into 2009, is from the *All Sky Automated Survey* ($ASAS$)-3 project of Pojmanski (2002). This automated observational program obtains one $V$ measurement per night, and their data base has 1221 points for MW Pav. With each observation, they provide a quality code of A through D, and we only used the highest quality A values. After visually removing a few additional points that had large deviations from the light curve, the final data set contained 836 $V$ observations. The only published radial velocities have been by Rucinski & Duerbeck (2006), and they acquired 18 and 13 measurements for the primary and secondary components, respectively, taken over three days in 1998. From this point forward, the four data sets are referred to as the Williamon, the Lapasset, the $ASAS$, and the RV data. Our objective was to solve simultaneously these data sets to improve parameter consistency (Wilson 1979; Van Hamme & Wilson 1984, 1985), but we quickly realized there were issues regarding the mass ratio, luminosities, temperatures, period, and starspots, which will be described throughout this section.

The light and velocity curve solutions were computed with the 2013 version of the Wilson-Devinney program. The WD program's physical model is described in detail in Wilson & Devinney (1971), Wilson (1979, 1990, 2012a,b), Van Hamme & Wilson (2007), and Wilson, Van Hamme & Terrell (2010). Mode 3 for overcontact binaries was used. The photometric observations in each data set were assigned a weight of 1, whereas the RV data were each given a weight of 10. Our curve-dependent weights were computed from the standard deviations that are listed in Table 3. Light level-dependent weights were applied inversely proportional to the square root of the light level. Gravity darkening ($g$) and bolometric albedo ($A$) coefficients were fixed at convective-envelope, canonical values from Lucy (1967) for both stars. We adopted a square-root limb darkening law with coefficients $x, y$ from Van Hamme (1993) for both components, and the detailed reflection treatment of Wilson (1990) was used with one reflection. Values of our non-varying parameters are listed in Table 4.

The first issue encountered was a discrepancy between the spectroscopic mass ratio of 0.22 (Rucinski & Duerbeck 2006) and the 0.18 photometric value by both Lapasset (1980) and our initial runs. The necessary adjustment to the photometric solution was the inclusion of a third light component. The luminosity terms and third light values for each photometric set are provided in Table 5. As a function of bandpass, the values are rather consistent among the three photometric sets. The third light star is brightest in the blue filter. From the Williamon data, $B - V = -0.222 \pm 0.052$ and $U - B = +0.074 \pm 0.047$. The corresponding Lapasset color indices are $-0.336 \pm 0.049$ and $+0.233 \pm 0.047$. Our derived mass ratio is $0.222 \pm 0.002$.

Houk & Cowley (1975) classified MW Pav as an F3 IV/V star. We set the primary's temperature at 6900 K, via the tables of Allen (2000), which was held fixed. We performed WD runs with the secondary's temperature allowed to vary and the result was a slightly warmer $6969 \pm 10$ K. Lu et al. (2007) comments it is often the case that W UMa systems have a hotter secondary than primary. Although we quote the WD error in the text and tables, we estimate the uncertainty in $T_1$ and $T_2$ to each be $\pm$ 200 K.

Given the 40-year range of photometry, it was not surprising that the orbital period value needed improvement. We reviewed the "O-C Gateway" website (BRNO 2011), but besides the times of minimum (TOM) published by Williamon (1971) and by Lapasset (1977, 1980), there were only two additional points, one each from *Hipparcos* (ESA 1997) and from Pojmanski (2002). These TOMs are provided in Table 6. Because of their non-uniform distribution during the date range, we used the WD program on all four of the data sets to calculate the parameters. We derived an epoch, period, and linear rate of change. The improved ephemeris is

$$\text{Light}_{\min} = \text{HJD}2440862.60793 \pm 0.00018 + 0.79498593 \pm 0.00000012\, E + (2.58 \pm 0.07) \times 10^{-10}\, E^2 \ .$$

The WD dP/dt term is $(6.50 \pm 0.19) \times 10^{-10}$.

The final issue concerned the need for starspots, and this included the questions of (a) which star or stars had a spot or spots, (b) whether the spots were hot or cool, and (c) whether the characteristics (location, size, and temperature) changed over time. The fit to the theoretical light curves was improved when a cool spot was added to the secondary's surface. This star was chosen instead of the primary because the bottom of secondary eclipse is flat but slanted due to the influence of the spot suddenly appearing at third contact. As we solved all of the data sets together, we utilized the WD program's ability to turn starspots on and off. This isolated the single spot during each of the three observational time ranges. However, the WD program can adjust at most two starspots during one run. First, we adjusted the Williamon and Lapasset spots and kept the $ASAS$ one unvarying. Once a solution was found, the pairs of spots adjusted were changed. We continued this iterative process until a solution was derived that gave the same results for the orbital elements no matter which spot pair was adjusted.

For the starspot, we held its latitude fixed on the equator. Attempts were made to determine a different latitude, but the WD software could not produce consistent results. Given that the binary system is seen nearly edge-on, it probably requires data of significantly higher precision to accurately determine a latitude position. The spot's longitude, angular radius, and temperature factor for each photometric set are listed in Table 5. It is noted that the longitude and radius changed little over time, whereas the temperature factor has been slightly increasing (i.e., getting warmer).

A simultaneous, base solution of the stellar parameters was derived and the orbital elements are given in Table 7. Absolute dimensions include masses of $M_1 = 1.514 \pm 0.063\, M_\odot$ and $M_2 = 0.327 \pm 0.014\, M_\odot$, and equal-volume radii of $R_1 = 2.412 \pm 0.034\, R_\odot$ and $R_2 = 1.277 \pm 0.019\, R_\odot$. The absolute dimensions are given in Table 8. Figure 1 shows the Williamon measurements along with the light curves computed in each bandpass from the orbital elements. The residuals to the fits are graphed in Figure 2. Likewise, Figures 3 and 4 show the Lapasset data, solution curves, and residuals. The $ASAS$ and RV results are presented in Figures 5 and 6, respectively.

The WD program provides geometrical information on the two stars. For overcontact binaries, relative radii are given in three directions: from the center toward the poles, toward the sides, and toward the back (i.e., away from the companion). In addition, it computed "equal-volume," mean radii ($<r>$) and the percentage of the Roche lobe ($<r>/<r>_{lobe}$) that is filled, which for both components is greater than 100%. The contact parameter or "fillout factor" $f$ (Van Hamme 1982b) is 60%. The relative radii are listed in Table 9. Figure 7 presents a series of images of the system from phases 0.55 to 0.95 to demonstrate how the cool spot distorts the light curves.

From the WD solution one obtains the bolometric magnitudes (see Table 8) and this information can be used to derive a distance. The primary's $M_{\rm bolo} = 2.069 \pm 0.130$ mag, and the bolometric correction from Flower (1996) for a sub-giant star with a temperature of 6900 K is $+0.028$ mag; therefore, $M_V = 2.042 \pm 0.130$ mag. The luminosity ratio is $0.297 \pm 0.004$ per the Williamon $V$ data. Thus, the primary is $0.282 \pm 0.003$ mag fainter than the combined $V$ mag. The "new *Hipparcos* reduction" by van Leeuwen (2007) listed a combined $V = 8.840 \pm 0.016$ mag. Adding the 0.282 mag and 8.840 mag values gives the primary's $V = 9.122 \pm 0.017$ mag. From the absolute and apparent magnitudes, the computed distance is $261 \pm 16$ pc. The *Hipparcos* data for MW Pav (HIP 102508) gives a parallax of $0.00480'' \pm 0.00108''$ (ESA 1997) and the van Leeuwen (2007) value is $0.00862'' \pm 0.00065''$. These parallaxes correspond to distances of $208.3 \pm 46.9$ pc and $209.2 \pm 42.0$ pc, respectively. Our greater distance may be due to the effects of interstellar extinction or the uncertainty in the absolute magnitude of the primary star.

## 4. DISCUSSION OF PREVIOUS SOLUTIONS

Lapasset (1977) derived a period of $0.79498855 \pm 0.00000091$ days. He binned the data into 75 normal points and solved for the photometric elements via the Russell-Merrill method (Russell & Merrill 1952); their technique included the standard rectification equation. The solution indicated a mass ratio of 0.12. Lapasset (1980) redetermined a photometric solution using the WD program. His three light curves were solved simultaneously, and solutions with and without a hot spot were determined. There was little difference in the results, and general values were $i = 85°$, $T_1 = 7620$ K, $T_2 = 7565$ K, and a mass ratio of 0.182. Visual inspection of the theoretical light curve fit at phase 0.25 (his Figure 1) showed the theoretical curve was significantly below the data points; Lapasset suggested a hot spot was needed. He found the system to be 43.6% to 50.4% overcontact, depending on the solution.

Rucinski & Duerbeck (2006) computed a solution of their radial velocity measurements. The 18 data points fit the primary's velocity well, but there was some scatter in the secondary's 13 points. They noted that "$K_2$ might be systematically underestimated at the available resolution." Their mass ratio was $0.228 \pm 0.008$.

Deb & Singh (2011) utilized the WD program to analyze 62 eclipsing binaries with the *ASAS* photometry and previously published radial velocity mass ratios. For MW Pav, they initially used the 0.228 mass ratio by Rucinski & Duerbeck (2006) with 1221 $V$ data points. Because they "found that their light curve could not be fitted properly, especially the minima, using the spectroscopic mass ratio," they allowed that parameter to vary. Their final result was $0.200 \pm 0.013$. They also adjusted the bolometric albedos from 0.50 to 0.70, and they included a third light component. As did Lapasset, Deb & Singh noted MW Pav showed the O'Connell (1951) effect of uneven outside-eclipse brightness at phases 0.25 and 0.75, but they declined to incorporate any spots to improve the fit. A comparison of their solution's absolute elements with our results is given in Table 8.

## 5. SUMMARY

We analyzed the southern A-type W UMa binary MW Pav using three sets of $UBV$ photometric observations and one set of previously published radial velocities. We determined the orbital elements and absolute dimensions with the Wilson-Devinney program and these are shown in Tables 5, 7, and 8. The best simultaneous fit to all the data requires that the system is in an overcontact configuration, that the secondary component has a large cool spot on its surface, and that there is a third light component. The mass ratio is $0.222 \pm 0.002$, and the individual masses are $M_1 = 1.514 \pm 0.063\, M_\odot$ and $M_2 = 0.327 \pm 0.014\, M_\odot$. The mean radii are $R_1 = 2.412 \pm 0.034\, R_\odot$ and $R_2 = 1.277 \pm 0.019\, R_\odot$. The stars have temperatures of 6900 K and $6969 \pm 10$ K, respectively. We recommend that times of minimum be monitored regularly, for the period has changed over 40 years. Whether the third light object is a member of the system has not been determined.

We thank Walter Van Hamme for valuable discussions about this system and about the WD program's various modes. In addition, he graciously computed the error-bars of the parameters in the final solution. This research made use of the SIMBAD database, operated at CDS, Strasbourg, France.

Table 1.   MW Pav Photometric Observations by Williamon[a]

| Helio. Julian Date (HJD − 2400000) | $\Delta V$ (mag) | Helio. Julian Date (HJD − 2400000) | $\Delta B$ (mag) | Helio. Julian Date (HJD − 2400000) | $\Delta U$ (mag) |
|---|---|---|---|---|---|
| 40862.5495 | −0.522 | 40862.5626 | −0.451 | 40862.5768 | −0.387 |
| 40862.5536 | −0.479 | 40862.5682 | −0.435 | 40862.5816 | −0.368 |
| 40862.5551 | −0.468 | 40862.5721 | −0.395 | 40862.5874 | −0.359 |
| 40862.5577 | −0.461 | 40862.5736 | −0.411 | 40862.5906 | −0.363 |
| 40862.5593 | −0.478 | 40862.5768 | −0.393 | 40862.5918 | −0.365 |

[a]Table 1 is presented in its entirety in the electronic edition of the PASP. A portion is shown here for guidance regarding its form and content.

Table 2.   MW Pav Photometric Observations by Lapasset[a]

| Helio. Julian Date (HJD − 2400000) | $\Delta V$ (mag) | $\Delta B$ (mag) | $\Delta U$ (mag) |
|---|---|---|---|
| 41587.5120 | +0.727 | +0.986 | +0.927 |
| 41587.5259 | +0.748 | +0.999 | +0.951 |
| 41587.5703 | +0.886 | +1.149 | +1.071 |
| 41587.5815 | +0.942 | +1.210 | +1.145 |
| 41587.5919 | +0.986 | +1.252 | +1.180 |

[a]Table 2 is presented in its entirety in the electronic edition of the PASP. A portion is shown here for guidance regarding its form and content.

Table 3. MW Pav Measurement Characteristics

| Curve | Observer | Data Points[a] | Normal Mag | $\sigma$ |
|---|---|---|---|---|
| $V$ | Williamon | 448 | $-0.8047$ | 0.012 |
| $B$ | Williamon | 448 | $-0.8286$ | 0.010 |
| $U$ | Williamon | 448 | $-0.9340$ | 0.014 |
| $V$ | Lapasset | 358 | $+0.6228$ | 0.010 |
| $B$ | Lapasset | 358 | $+0.8610$ | 0.011 |
| $U$ | Lapasset | 358 | $+0.8022$ | 0.012 |
| $V$ | $ASAS$ | 836 | $+8.6047$ | 0.013 |
| $RV_1$ | Rucinski & Duerbeck | 18 | $\cdots$ | 21 km s$^{-1}$ |
| $RV_2$ | Rucinski & Duerbeck | 13 | $\cdots$ | 29 km s$^{-1}$ |

[a]Per the discussion in Section 2, we excluded 43 values from Lapasset's data, taken on HJD 2441589, from the WD runs; however, all of the Lapasset observations are listed in Table 2.

Table 4.   Non-Varying WD Parameters

| Parameter | Symbol | Value |
|---|---|---|
| Rotation/Orbit Ratio | $F_1, F_2$ | $1.00, 1.00$ |
| Albedo (bolo) | $A_1, A_2$ | $0.50, 0.50$ |
| Gravity Darkening | $g_1, g_2$ | $0.32, 0.32$ |
| Limb Darkening (bolo) | $x_1, y_1$ | $+0.086, +0.638$ |
| Limb Darkening (bolo) | $x_2, y_2$ | $+0.086, +0.638$ |
| | | |
| Limb Darkening ($V$) | $x_1, y_1$ | $+0.063, +0.724$ |
| Limb Darkening ($V$) | $x_2, y_2$ | $+0.063, +0.724$ |
| Limb Darkening ($B$) | $x_1, y_1$ | $+0.191, +0.691$ |
| Limb Darkening ($B$) | $x_2, y_2$ | $+0.191, +0.691$ |
| Limb Darkening ($U$) | $x_1, y_1$ | $+0.088, +0.817$ |
| Limb Darkening ($U$) | $x_2, y_2$ | $+0.088, +0.817$ |

Table 5. MW Pav Data Set Characterisitics

| Parameter | Symbol | Williamon | Lapasset | $ASAS$ |
|---|---|---|---|---|
| Luminosity ratio $(V)$ | $L_1/(L_1+L_2)$ | 0.7712 ± 0.0149 | 0.7712 ± 0.0148 | 0.7712 ± 0.0149 |
| Luminosity ratio $(B)$ | $L_1/(L_1+L_2)$ | 0.7684 ± 0.0155 | 0.7684 ± 0.0160 | |
| Luminosity ratio $(U)$ | $L_1/(L_1+L_2)$ | 0.7707 ± 0.0161 | 0.7707 ± 0.0157 | |
| Third Light $(V)$[a] | $l_3/(l_1+l_2+l_3)$ | 0.1012 ± 0.0075 | 0.1025 ± 0.0075 | 0.1167 ± 0.0074 |
| Third Light $(B)$[a] | $l_3/(l_1+l_2+l_3)$ | 0.1242 ± 0.0074 | 0.1396 ± 0.0076 | |
| Third Light $(U)$[a] | $l_3/(l_1+l_2+l_3)$ | 0.1161 ± 0.0079 | 0.1126 ± 0.0077 | |
| Spot Parameters | | | | |
| Latitude (deg) | $\text{Lat}_{spot}$ | 0.0[b] | 0.0[b] | 0.0[b] |
| Longitude (deg) | $\text{Long}_{spot}$ | 293.0 ± 2.3 | 287.1 ± 2.3 | 285.4 ± 4.3 |
| Radius (deg) | $\text{R}_{spot}$ | 28.3 ± 1.9 | 27.9 ± 4.7 | 30.9 ± 11.1 |
| Temperature Factor | $\text{T}_{spot}$ | 0.687 ± 0.070 | 0.788 ± 0.096 | 0.831 ± 0.143 |

[a]The Third Light parameters are in units of total light at phase 0.25.

[b]Adopted value, see Section 3 in the text.

Table 6. Times of Primary Minima for MW Pav

| HJD − 2400000 | Filter | Observer |
|---|---|---|
| 40862.6100 | V | Williamon |
| 40862.6080 | B | Williamon |
| 40862.6111 | U | Williamon |
| 40870.5615 | V | Williamon |
| 40870.5584 | B | Williamon |
| 40870.5585 | U | Williamon |
| 41587.6352 | V | Lapasset |
| 41587.6347 | B | Lapasset |
| 41587.6335 | U | Lapasset |
| 41606.7144 | V | Lapasset |
| 41606.7132 | B | Lapasset |
| 41606.7126 | U | Lapasset |
| 48500.0630 | V | *Hipparcos* |
| 51874.7970 | V | Pojmanski |

Table 7.  Light and Velocity Curve Results for MW Pav[a]

| Parameter | Symbol | Value |
|---|---|---|
| Inclination (deg) | $i$ | 86.39 ± 0.63 |
| Mass ratio | $M_2/M_1$ | 0.222 ± 0.002 |
| Surface potential | $\Omega_1$ | 2.185 ± 0.005 |
| Surface potential | $\Omega_2$ | 2.185[b] |
| Temperature (K) | $T_1$ | 6900[c] |
| Temperature (K) | $T_2$ | 6969 ± 10 |
| Eccentricity | $e$ | 0.0[c] |
| Systemic velocity (km s$^{-1}$) | $\gamma$ | −41.37 ± 1.33 |
| Semimajor axis ($R_\odot$) | $a$ | 4.427 ± 0.048 |
| Epoch (HJD) | $T_o$ | 2,440,862.60793 ± 0.00018 |
| Period (d) | $P$ | 0.79498593 ± 0.00000012 |
| First Derivative of Period Change | $dP/dt$ | (6.50 ± 0.19) × 10$^{-10}$ |

[a]Wilson-Devinney simultaneous solution, including proximity and eclipse effects, of the light and velocity data.

[b]Set equal to the surface potential of the primary.

[c]Adopted value, see Section 3 in the text.

Table 8. Fundamental Parameters of MW Pav

| Parameter | This Study | Deb & Singh |
|---|---|---|
| Primary | | |
| $M\,(M_\odot)$ | 1.514 ± 0.063 | 1.520 ± 0.045 |
| $R\,(R_\odot)$ | 2.412 ± 0.034 | 2.456 ± 0.033 |
| $L/L_\odot$ | 11.819 ± 1.409 | 12.118 ± 1.449 |
| $M_{bol}$ (mag) | 2.069 ± 0.130 | |
| $\log g$ (cm s$^{-2}$) | 3.854 ± 0.007 | |
| | | |
| Secondary | | |
| $M\,(M_\odot)$ | 0.327 ± 0.014 | 0.346 ± 0.019 |
| $R\,(R_\odot)$ | 1.277 ± 0.019 | 1.277 ± 0.019 |
| $L/L_\odot$ | 3.314 ± 0.396 | 3.193 ± 0.390 |
| $M_{bol}$ (mag) | 3.459 ± 0.130 | |
| $\log g$ (cm s$^{-2}$) | 3.740 ± 0.008 | |
| | | |
| System | | |
| $i$ (deg) | 86.39 ± 0.63 | 84.81 ± 0.60 |
| $M_2/M_1$ | 0.222 ± 0.002 | 0.200 ± 0.013 |
| $f$ | 0.60 | 0.52 |
| $T_1$ (K) | 6900 | 6881 ± 160 |
| $T_2$ (K) | 6969 ± 10 | 6837 ± 158 |

Table 9. Model Radii for MW Pav

| Parameter | Value |
|---|---|
| $r_1$ (pole) | $0.5006 \pm 0.0010$ |
| $r_1$ (side) | $0.5504 \pm 0.0013$ |
| $r_1$ (back) | $0.5803 \pm 0.0012$ |
| $<r_1>$[a] | $0.5448 \pm 0.0010$ |
| $<r_1>/<r_1>_{\rm lobe}$ | $1.0543 \pm 0.0031$ |
| $r_2$ (pole) | $0.2614 \pm 0.0067$ |
| $r_2$ (side) | $0.2756 \pm 0.0085$ |
| $r_2$ (back) | $0.3370 \pm 0.0242$ |
| $<r_2>$[a] | $0.2885 \pm 0.0016$ |
| $<r_2>/<r_2>_{\rm lobe}$ | $1.1265 \pm 0.0072$ |

[a] "Equal-volume" mean radii.

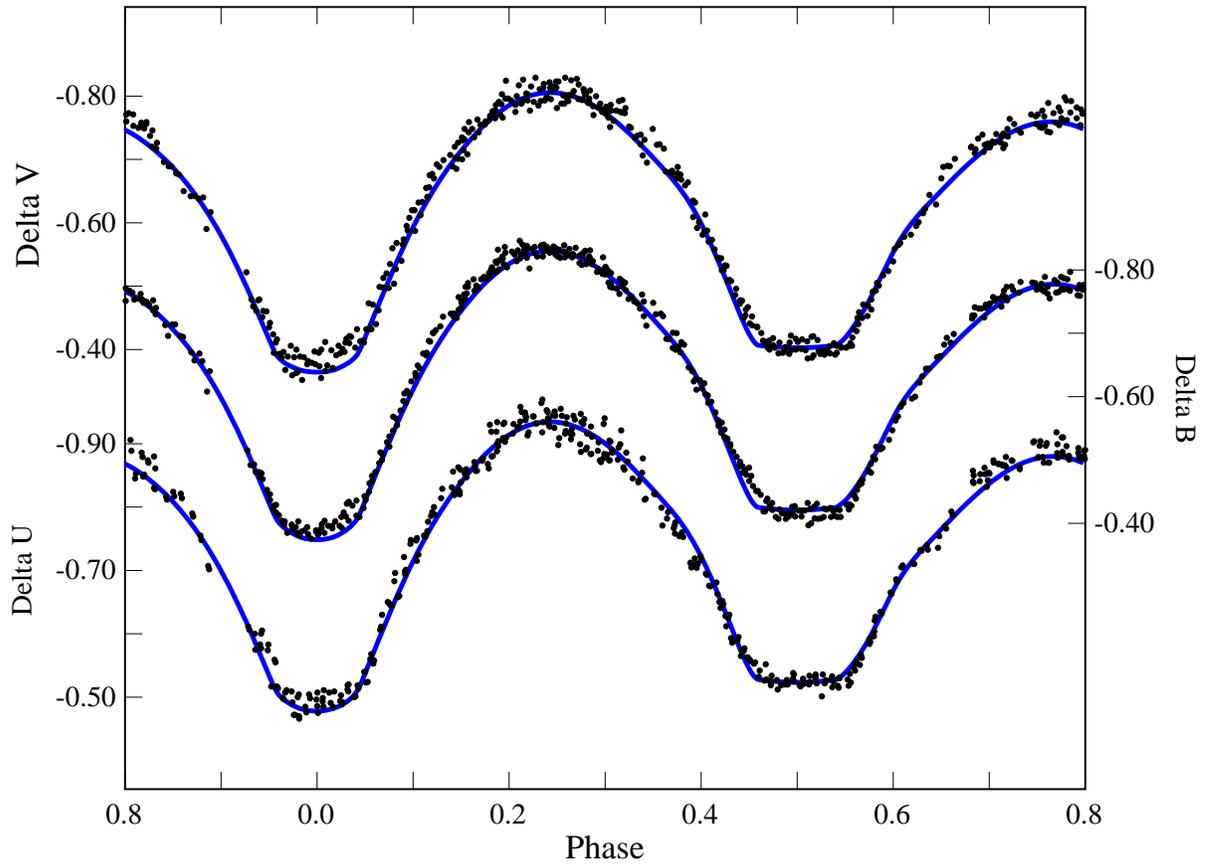

Fig. 1.— The Williamon differential $UBV$ magnitudes of MW Pav are plotted with the Wilson-Devinney solution curves based on the three photometric sets and the RV measurements. The system is an overcontact binary, and a large, cool spot is on the secondary, which depresses the light curve in the 0.80 phase region (see Table 5).

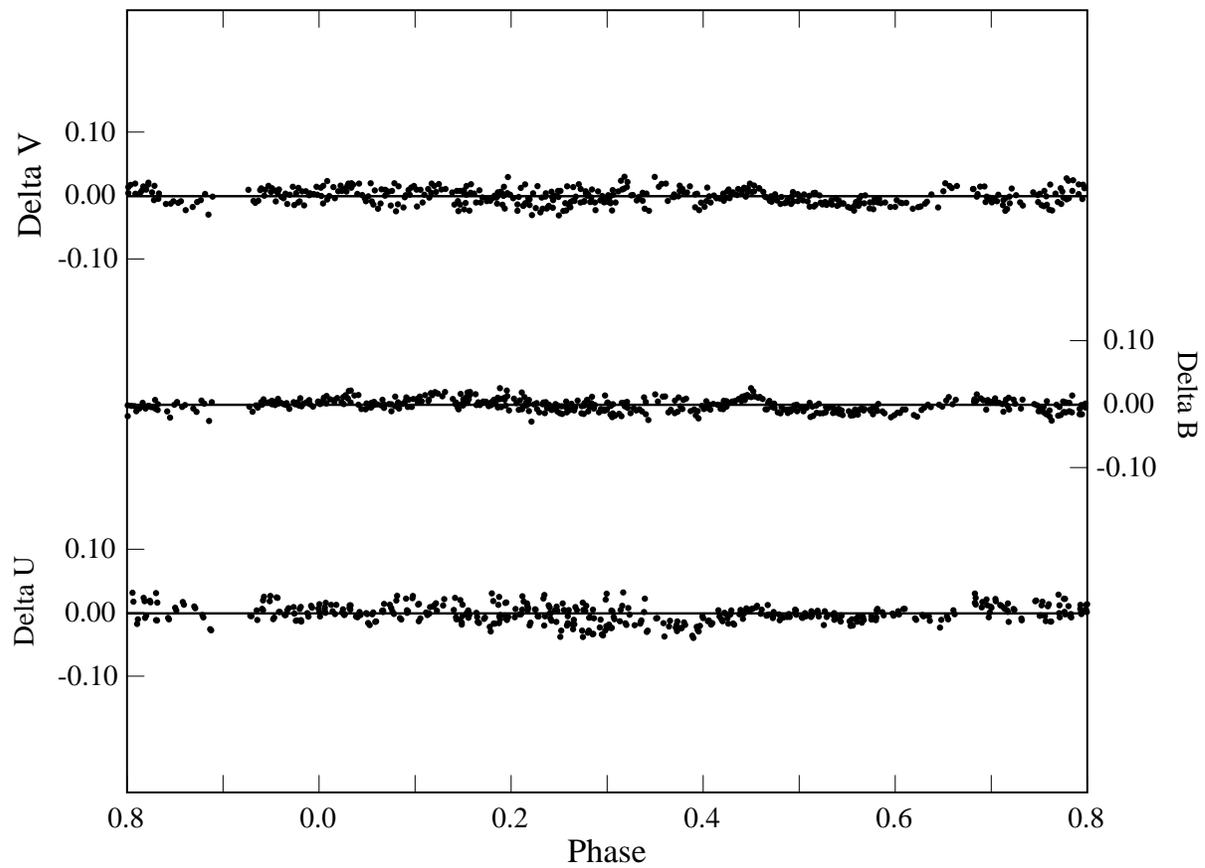

Fig. 2.— Residuals to the fit of the Williamon $UBV$ photometry provided by the solution light curves.

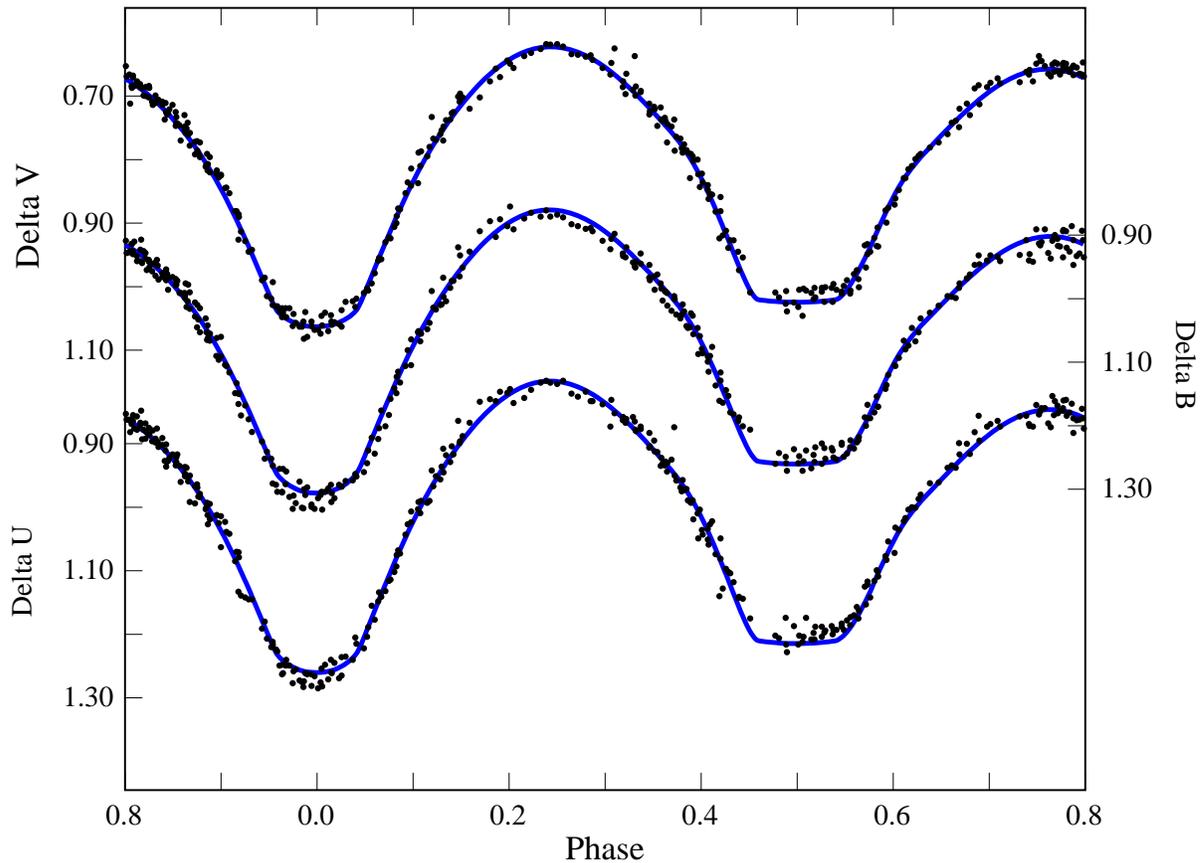

Fig. 3.— The Lapasset differential $UBV$ magnitudes of MW Pav are plotted with the Wilson-Devinney solution curves based on the three photometric sets and the RV measurements. The cool spot has a different size, temperature, and position than that of the Williamon data solution, although still near phase 0.80. The spot characteristics are given in Table 5.

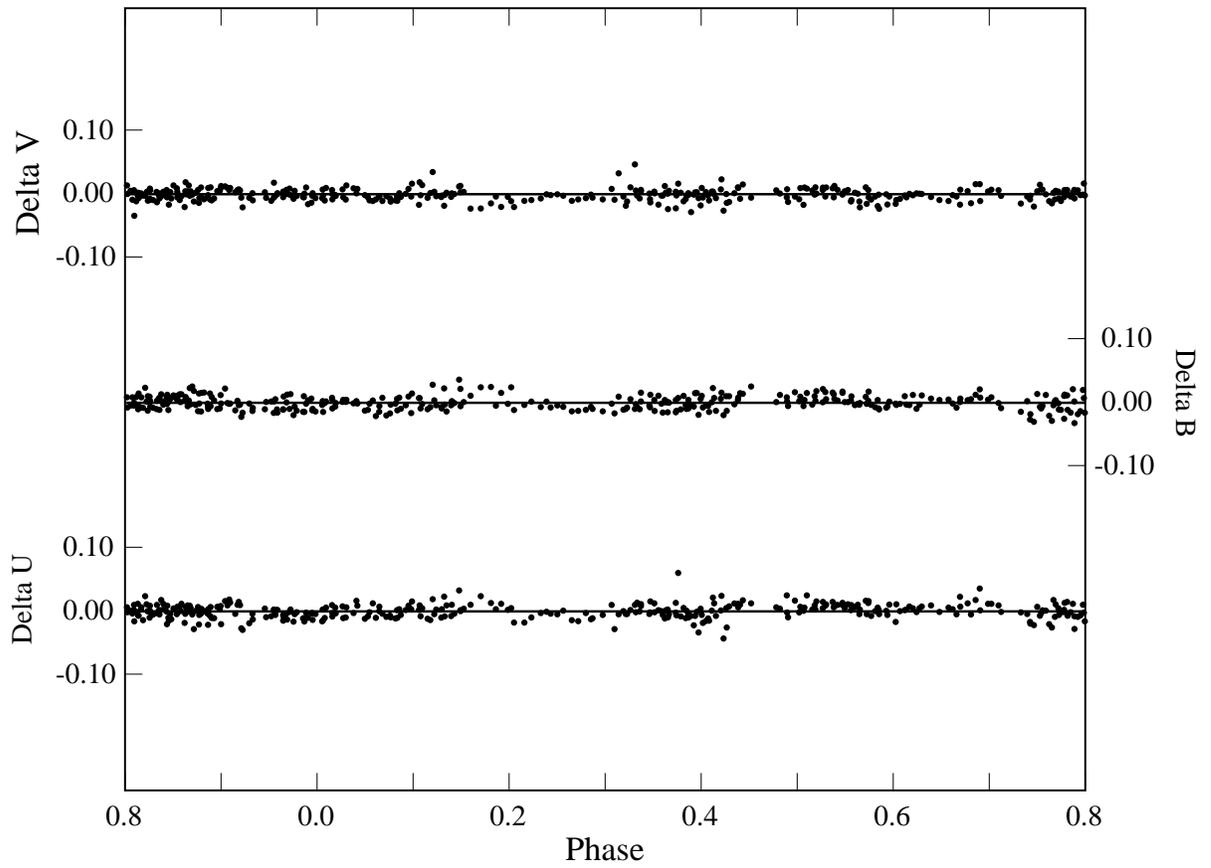

Fig. 4.— Residuals to the fit of the Lapasset $UBV$ photometry provided by the solution light curves.

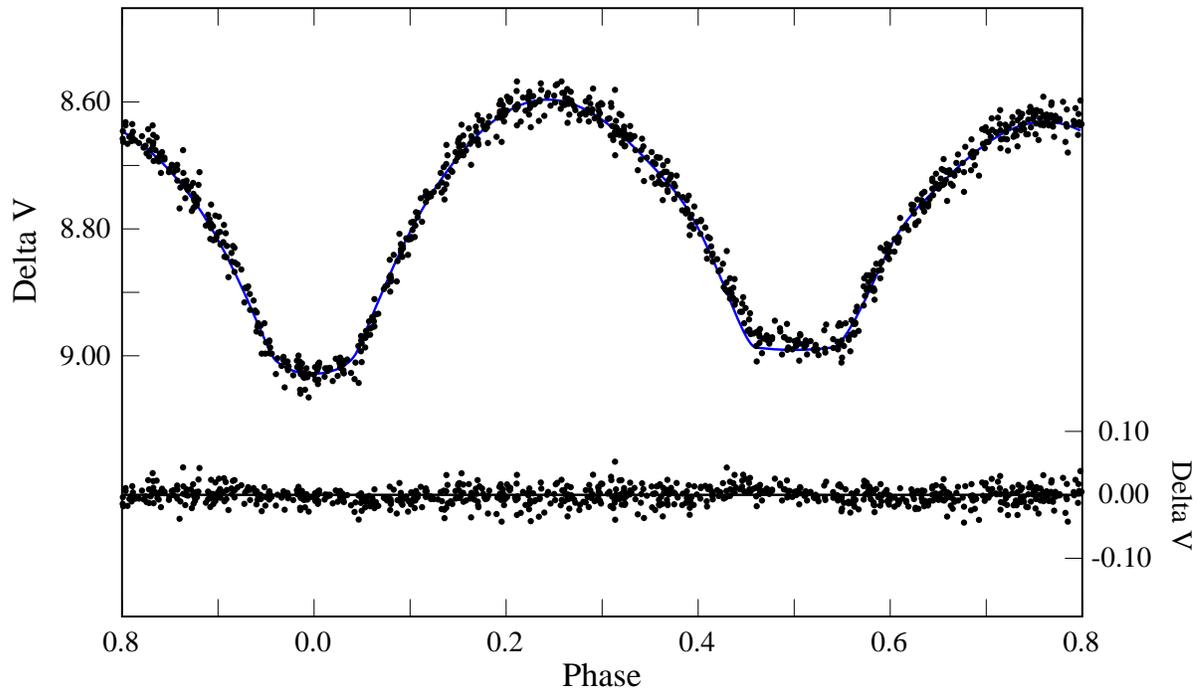

Fig. 5.— The *ASAS* differential *UBV* magnitudes (Pojmanski 2002) of MW Pav are plotted with the Wilson-Devinney solution curves based on the three photometric sets and the RV measurements. The cool spot has a different size, temperature, and position than that of the Williamon and Lapasset data solutions, but it continues to reside near phase 0.80. The spot characteristics are given in Table 5. Residuals to the fit, provided by the solution light curve, are plotted at the bottom of the figure.

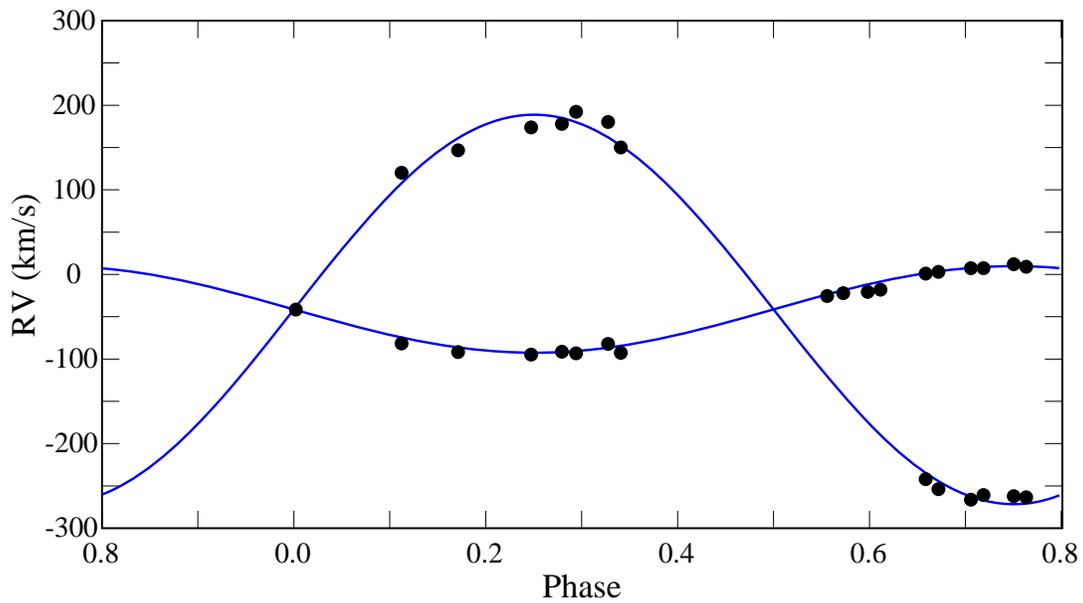

Fig. 6.— The radial velocities obtained by Rucinski & Duerbeck (2006) of MW Pav are plotted with the Wilson-Devinney solution curves for the combined $UBV$ and RV data. Zero phase is at the time of primary eclipse.

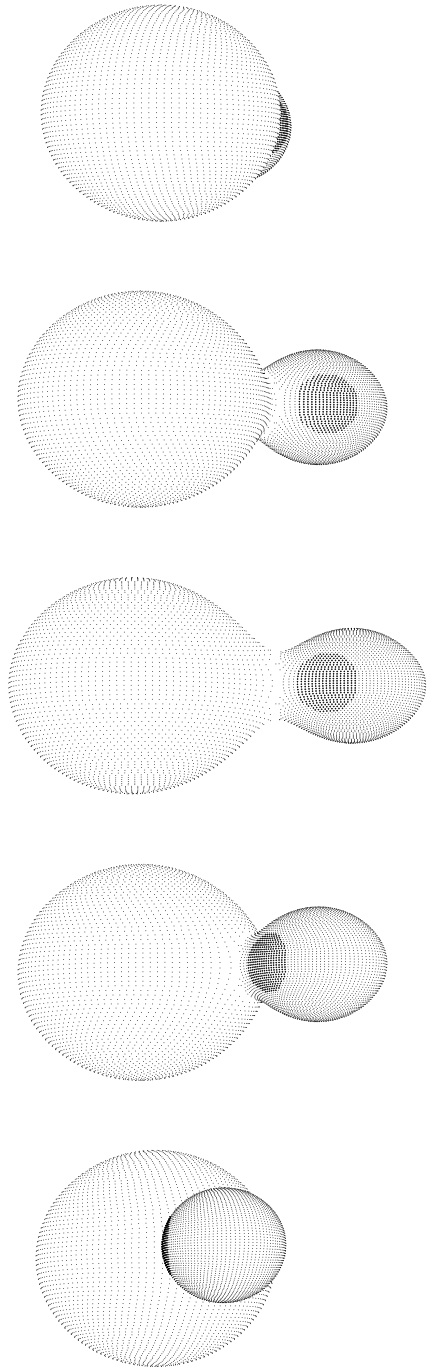

Fig. 7.— The distortion of the light curve by a cool spot is demonstrated via the images of MW Pav at phases 0.55, 0.65, 0.75, 0.85, and 0.95. The Williamon starspot is modeled, and its characteristics are given in Table 5. The system is an A-type, W UMa binary, so both stars have overfilled their Roche lobes.

25